\documentclass[]{spie}  

\usepackage{amsmath,amsfonts,amssymb}
\usepackage{graphicx}
\usepackage[colorlinks=true, allcolors=blue]{hyperref}
\usepackage[separate-uncertainty=true]{siunitx}
\DeclareSIUnit\pixel{px}
\usepackage[version=4]{mhchem}
\usepackage{multirow}
\usepackage{cleveref}
\usepackage{placeins}

\title{Noise and dose reduction in CT brain perfusion acquisition by projecting time attenuation curves onto lower dimensional spaces}

\author[a]{Vojt\v{e}ch~Kulvait}
\author[b]{Philip~Hoelter}
\author[b]{Arnd~Doerfler}
\author[a]{Georg~Rose}
\affil[a]{Institute for Medical Engineering and Research Campus STIMULATE, University of Magdeburg, Magdeburg, Germany}
\affil[b]{Department of Neuroradiology, University Hospital Erlangen, Friedrich-Alexander-Universit\"{a}t (FAU) Erlangen-N\"{u}rnberg, Erlangen, Germany
}

\authorinfo{Send correspondence to: kulvait@gmail.com}

\begin{document} 
\maketitle

\begin{abstract}
CT perfusion imaging (CTP) plays an important role in decision making for the treatment of acute ischemic stroke with large vessel occlusion. Since the CT perfusion scan time is approximately one minute, the patient is exposed to a non-negligible dose of ionizing radiation. However, further dose reduction increases the level of noise in the data and the resulting perfusion maps. We present a method for reducing noise in perfusion data based on dimension reduction of time attenuation curves. For dimension reduction, we use either the fit of the first five terms of the trigonometric polynomial or the first five terms of the SVD decomposition of the time attenuation profiles. CTP data from four patients with large vessel occlusion and three control subjects were studied. To compare the noise level in the perfusion maps, we use the wavelet estimation of the noise standard deviation implemented in the scikit-image package. We show that both methods significantly reduce noise in the data while preserving important information about the perfusion deficits. These methods can be used to further reduce the dose in CT perfusion protocols or in perfusion studies using C-arm CT, which are burdened by high noise levels.
\end{abstract}


\section{INTRODUCTION}
\label{sec:introduction}  

CT perfusion imaging (CTP) helps select population of patients with acute ischemic stroke and large vessel occlusion within 6 to 24 hours from the symptoms onset who are suitable for mechanical thrombectomy \cite{Albers2018,powers2019guidelines}. A typical perfusion scan takes approximately one minute and involves a significant dose of radiation for the patient. Low dose protocols cause additional noise in the data and resulting perfusion maps. Dimension reduction of the temporal attenuation profiles is a tool that can reduce noise in the data, as has been shown in the case of C-arm CT perfusion studies \cite{Bannasch2018, Haseljic2021}. In this paper, we apply dimension reduction techniques to CT perfusion data to decrease noise with applications to radiation dose reduction.

\section{METHODS}
\begin{figure}[t]
\centering
    \includegraphics[width=\textwidth]{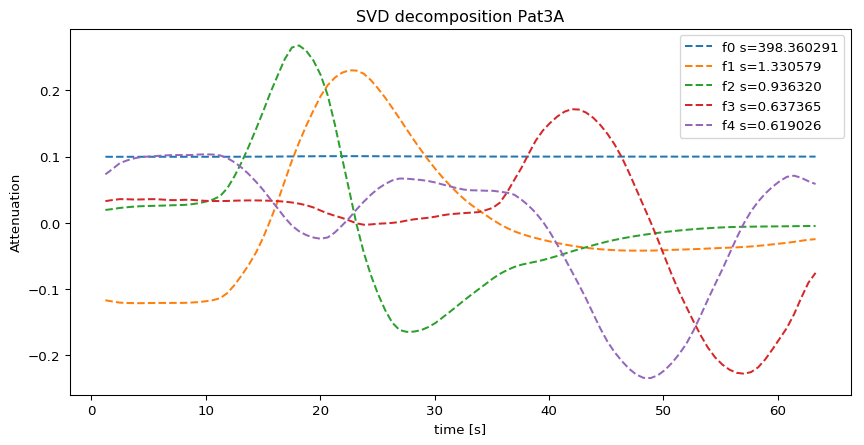}
    \caption{Example SVD basis derived from the patient data. In the legend, there is a singular value corresponding to each temporal vector.}
    \label{fig:svd}
\end{figure}

CT perfusion data from 4 acute ischemic stroke patients and 3 control subjects were used. Time attenuation curves (TACs) were obtained from individual CTP acquisitions. For noise reduction we used two different dimension reduction techniques. In the first approach, we fit the first 5 terms of the trigonometric polynomial including the constant to the TACs. In the second approach, we first perform an SVD decomposition of the TACs, see Fig.~\ref{fig:svd}. We then fit the 5 profiles corresponding to the highest singular values to the data. A Gaussian blurr with $\sigma=\SI{2}{\pixel}$ was added into the original volume data and the reduced-dimension data sets and these data were processed using a standard deconvolution-based perfusion processing method \cite{Fieselmann2011}. Perfusion maps of CBF, CBV, MTT and TTP were generated. To compare the noise levels in the perfusion maps, we use a wavelet-based estimator \texttt{skimage.restoration.estimate\_sigma} of the noise standard deviation implemented in the scikit-image package. 

\section{RESULTS}
\label{sec:results}
\begin{figure}[b]
\centering
    \includegraphics[width=\textwidth]{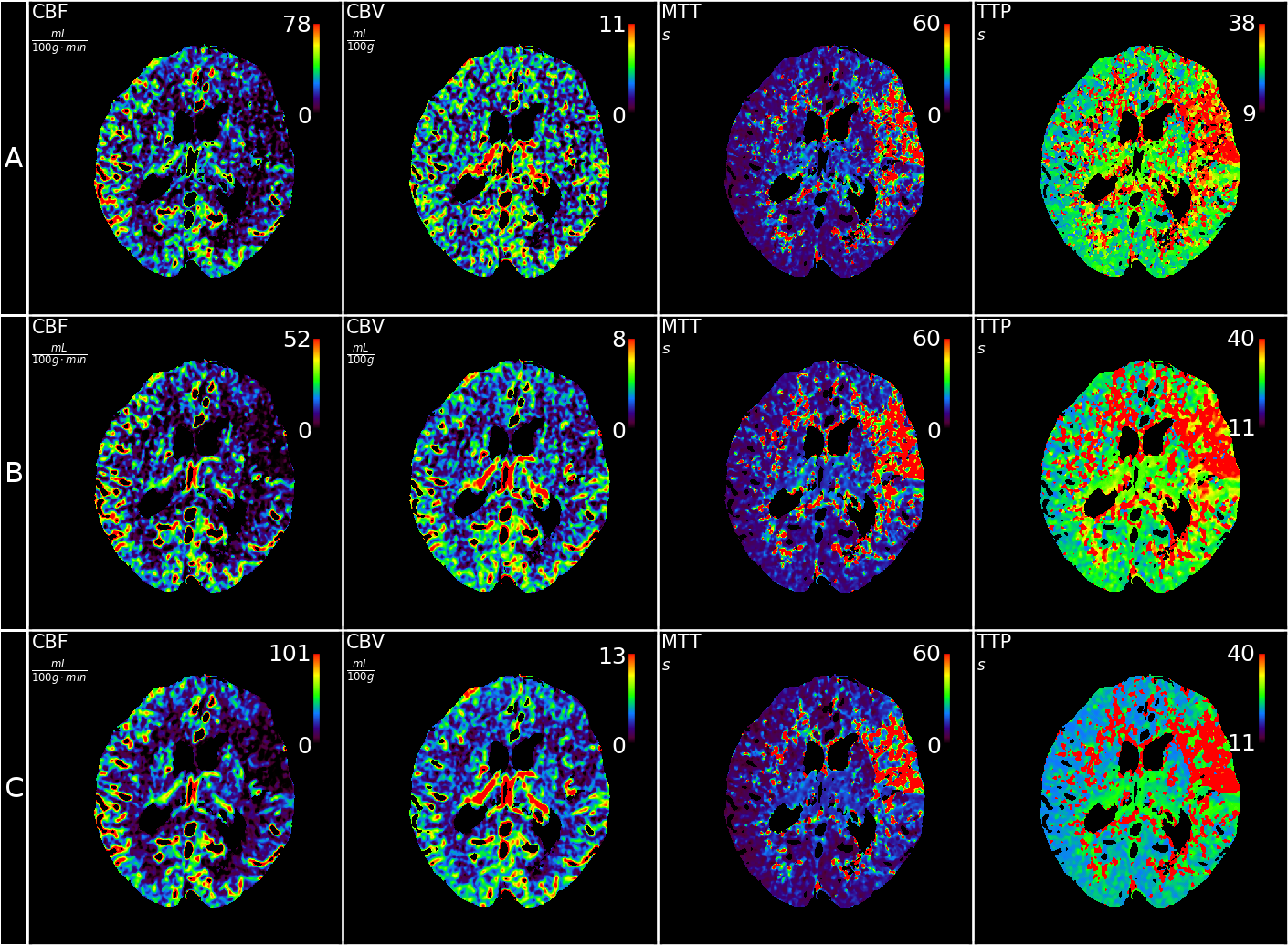}
    \caption{Comparison of the original data (A) with the data of reduced dimension using trigonometric functions (B) or SVD decomposition of temporal profiles (C). }
    \label{fig:perfusion}
\end{figure}

Tables \ref{tab:correlation_tgn} and \ref{tab:correlation_svd} summarize Pearson correlation coefficients when comparing data with and without dimension reduction. Table \ref{tab:correlation_tgn} contains the correlation coefficients for dimension reduction using trigonometric functions and table \ref{tab:correlation_svd} contains the correlation coefficients for dimension reduction using SVD decomposition. Figure \ref{fig:perfusion} shows the differences between data processed with and without dimension reduction on the example of a patient with large vessel occlusion. Table \ref{tab:sigma} summarizes the estimates of the standard deviation of the noise using the \texttt{skimage.restoration.estimate\_sigma} function of the scikit-image package. Paired T-test showed a significant reduction in noise level with $p<0.01$ values for all perfusion parameters and for both dimension reduction methods. 

\begin{table}[b]
\begin{center}
\begin{tabular}{l|llll|lll}
\multicolumn{1}{l}{}&\multicolumn{4}{c}{Patient}&\multicolumn{3}{c}{Control}\\
& 1A & 2A & 3A & 4A & C1 & C2 & C3 \\
\hline
CBF & 0.96   & 0.92  &  0.96  &  0.97  & 0.96  &  0.97  &  0.99  \\
CBV & 0.94   & 0.90  &  0.92  &  0.92  & 0.93  &  0.95  &  0.97  \\
MTT & 0.81   & 0.78  &  0.79  &  0.78  & 0.76  &  0.73  &  0.74  \\
TTP & 0.58   & 0.59  &  0.59  &  0.58  & 0.56  &  0.39  &  0.49  \\
\end{tabular}
\end{center}
\caption{Perarson correlation coefficients comparing data without dimension reduction with data of reduced dimension using  first five terms of trigonometric polynomial.}
\label{tab:correlation_tgn}
\end{table}

\begin{table}[b]
\begin{center}
\begin{tabular}{l|llll|lll}
\multicolumn{1}{l}{}&\multicolumn{4}{c}{Patient}&\multicolumn{3}{c}{Control}\\
& 1A & 2A & 3A & 4A & C1 & C2 & C3 \\
\hline
CBF & 0.97   & 0.97  &  0.95  &  0.95  & 0.96  &  0.98  &  0.98  \\
CBV & 0.96   & 0.96  &  0.92  &  0.92  & 0.96  &  0.97  &  0.97  \\
MTT & 0.82   & 0.79  &  0.73  &  0.75  & 0.75  &  0.82  &  0.74  \\
TTP & 0.59   & 0.70  &  0.60  &  0.55  & 0.59  &  0.49  &  0.51  \\
\end{tabular}
\end{center}
\caption{Perarson correlation coefficients comparing data without dimension reduction with data of reduced dimension using  first five terms of SVD decomposition.}
\label{tab:correlation_svd}
\end{table}

%
\begin{table}[t]
\begin{center}
\begin{tabular}{c|SSS}
    & \textbf{No dimension reduction}      & \textbf{Trigonometric functions}          & \textbf{SVD decomposition}                \\
    \hline
CBF & $\num{0.21(1)}$   & $\num{0.12(1)}$  & $\num{0.14(1)}$   \\
CBV & $\num{0.033(1)}$ & $\num{0.016(1)}$ & $\num{0.021(1)}$ \\
MTT & $\num{0.19(1)}$   & $\num{0.13(1)}$   & $\num{0.10(1)}$   \\
 TTP & $\num{0.99(7)}$     & $\num{0.31(1)}$     & $\num{0.30(1)}$    
\end{tabular}
\end{center}
\caption{Noise standard deviation estimates using the \texttt{skimage.restoration.estimate\_sigma} function of the scikit-image package for data without dimension reduction and data with dimension reduction using trigonometric functions and SVD decomposition.}
\label{tab:sigma}
\end{table}



\section{CONCLUSIONS}
\label{sec:conclusion}

Table \ref{tab:sigma} shows that both dimension reduction methods significantly reduce the noise in the perfusion maps. Dimension reduction using trigonometric polynomial and SVD decomposition had very comparable effects on the data used, see \cref{tab:correlation_tgn,tab:correlation_svd,tab:sigma}. The image data, see Figure \ref{fig:perfusion}, show that the noise reduction using both dimension reduction methods is most significant for the CBV map. The data suggest that SVD reduction can stabilize the TTP and MTT maps better. 


Both dimension reduction methods preserve well the essential patterns of perfusion deficits in perfusion maps as seen, for example, in Figure \ref{fig:perfusion}. Our proposed method of noise reduction in CT perfusion data can be used to further reduce the dose in CT perfusion acquisitions. We will also investigate these dimension reduction methods to process interventional perfusion data from C-arm CT systems.

\acknowledgments 
 
The work of this paper is partly funded by the Federal Ministry of Education and Research within the Research Campus STIMULATE under the number 13GW0473A.

\bibliographystyle{spiebib} 
\bibliography{bibliography.bib} 
\textit{This work has not been submitted, published or presented elsewhere.}

\end{document}